# Hyperspectral Super-Resolution: A Coupled Tensor Factorization Approach

Charilaos I. Kanatsoulis, *Student Member, IEEE,* Xiao Fu, *Member, IEEE,*
Nicholas D. Sidiropoulos, *Fellow, IEEE,* and Wing-Kin Ma, *Fellow, IEEE*

*Abstract*—Hyperspectral super-resolution refers to the problem of fusing a hyperspectral image (HSI) and a multispectral image (MSI) to produce a super-resolution image (SRI) that has fine spatial and spectral resolution. State-of-the-art methods approach the problem via low-rank matrix approximations to the matricized HSI and MSI. These methods are effective to some extent, but a number of challenges remain. First, HSIs and MSIs are naturally third-order tensors (data "cubes") and thus matricization is prone to loss of structural information—which could degrade performance. Second, it is unclear whether or not these low-rank matrix-based fusion strategies can guarantee identifiability or exact recovery of the SRI. However, identifiability plays a pivotal role in estimation problems and usually has a significant impact on performance in practice. Third, the majority of the existing methods assume that there are known (or easily estimated) degradation operators applied to the SRI to form the corresponding HSI and MSI—which is hardly the case in practice. In this work, we propose to tackle the super-resolution problem from a tensor perspective. Specifically, we utilize the multidimensional structure of the HSI and MSI to propose a coupled tensor factorization framework that can effectively overcome the aforementioned issues. The proposed approach guarantees the identifiability of the SRI under mild and realistic conditions. Furthermore, it works with little knowledge of the degradation operators, which is clearly an advantage over the existing methods. Semi-real numerical experiments are included to show the effectiveness of the proposed approach.

## I. Introduction

IMAGE fusion from multiple sensors has attracted much attention from several communities, including signal and image processing, remote sensing, and computer vision, because it is very useful in a lot of applications [1]–[3]. Recently, the remote sensing community has invested significant effort in exploring hyperspectral and multispectral image fusion. This problem is known as *hyperspectral super-resolution* (HSR) or *hyperspectral-multispectral fusion* [4]. The ultimate goal of HSR is to integrate information from a hyperspectral image (HSI), which features high spectral resolution but relatively coarse spatial resolution, and a (co-registered) multispectral image (MSI), which has fine spatial resolution but low spectral resolution, to produce a super-resolution image (SRI) that has high resolution in both spatial and spectral domains. This task is very well-motivated, since an SRI is of great interest in multiple analytical tasks – e.g., small object tracking and identification. However, it is considered very costly to improve both the spectral and the spatial resolution of the multiband sensors simultaneously, due to key hardware limitations [4]. Nevertheless, HSR techniques allow the construction of an SRI via fusing images that are captured by existing sensors [5], [6].

HSR is a long-standing problem in remote sensing. For example, a lot of early works in the 1990s and 2000s addressed the problem of hyperspectral pansharpening, which fuses an HSI and a panchromatic image to produce an SRI. This problem has a similar flavor as HSR; see a comprehensive review in [6], [7]. Existing pansharpening methods include component substitution (CS) [8], [9] and multiresolution analysis (MRA) [10]–[12], which work by injecting details from the panchromatic image into the HSI. Attempts have been made to use pansharpening type methods for HSR. They are mainly based on wavelet techniques [13], [14] or try to generalize CS and MRA pansharpening algorithms for HSR purposes [15]. However, such methods were found to have limitations with enhancing the spatial resolution of every hyperspectral band in practice [4].

Over the past few years, there has been renewed interest for HSR, which is largely triggered by the advances in modern optimization and matrix factorization techniques. Numerous recent methods for HSR utilize low-rank matrix factorization models [16]–[24]. The idea is to take advantage of the low-rank matrix structure of the matricized HSI and MSI. The low-rank structure stems from the so-called linear mixure model (LMM) that is widely employed for modeling hyperspectral/multispectral images. Under LMM, every spectral pixel of the HSI or MSI is modeled as a convex combination of the spectral signatures of several materials (or *endmembers*). This representation is physically intuitive and has enabled developments in another research topic, called hyperspectral unmixing [25]–[29]. More importantly, under LMM, all the pixels reside in a low-dimensional subspace spanned by a number of endmembers—which makes the matricized HSI/MSI of low rank. Several HSR approaches work under this model. For example, the works in [17], [20], [23] perform (coupled) low-rank factorization of the matricized HSI and MSI to estimate the spectral signatures of the endmembers (from HSI) and the corresponding high-resolution spatial distribution of the pixels (from MSI). Then the SRI is constructed by combining these

C.I. Kanatsoulis is with the Department of ECE, University of Minnesota, Minneapolis, MN 55455, USA; email: kanat003@umn.edu. X. Fu is with the School of EECS, Oregon State University, Corvallis, OR 97331, USA; email: xiao.fu@oregonstate.edu. N. D. Sidiropoulos is with the Department of ECE, University of Virginia, Charlottesville, VA 22904, USA; email: nikos@virginia.edu. W. K. Ma is with the Department of EE, The Chinese University of Hong Kong, Shatin, N.T., Hong Kong; email: wkma@cuhk.edu.hk.

two estimated matrices. A number of variants exist [16], [19], [21], [22], using different data representations and algorithms, all building upon low-rank modeling of the matricized HSI and MSI.

The matrix factorization approaches are effective to some extent, but several challenges remain. First, multiband images are naturally third-order tensors (i.e., data cubes whose elements are indexed by three indices). However, the low-rank matrix factorization-based approaches ignore such structure since these methods all reshape the 3D images to 2D matrices, and discard tensorial structure in the sequel. Consequently, critical dependence information across the three dimensions is ignored—which makes it hard to further improve the matrix factorization-based approaches by utilizing such dependence. Second, it is unclear whether the matrix-based HSI-MSI fusion criteria proposed in [16]–[20] can guarantee identifiability of the ground-truth SRI; i.e., there is no theoretical support for the identifiability of these methods. However, identifiability is known to be essential in such estimation problems in signal processing, since it asserts the soundness of the criteria and affects performance in practice in a significant way [30]–[34]. Whereas identifiability is usually conceptualized under a noiseless setting, lack of identifiability means that there are many equivalent possibilities for the ground truth *even if one fixes the modeling residuals*. Recall linear least squares to see this: a noisy under-determined linear model $\mathbf{y} = \mathbf{Ax} + \mathbf{w}$ (fat full row-rank $\mathbf{A}$) admits many solutions for $\mathbf{x}$ even if we fix (know) the correct $\mathbf{w}$, as with $\tilde{\mathbf{y}} := \mathbf{y} - \mathbf{w}$, $\tilde{\mathbf{y}} = \mathbf{Ax}$ admits many solutions. Third, the majority of existing approaches assume that the degradation operators from the virtual SRI to HSI and MSI are known [17]–[20], or that such operators can be easily estimated [16], [35]. This assumption is considered rather restrictive, since the spatial degradation is not a simple sum; it involves blurring, downsampling, and jitter.

**Contributions:** In this work, we propose a novel hyperspectral super-resolution approach. Our approach starts with the fact that both HSI and MSI images are space-space-spectrum "cubes", and thus can be naturally represented as third-order tensors [30]. Tensors have a number of favorable properties that matrices do not have. For example, tensors admit canonical polyadic decomposition (CPD), which captures dependencies across the different dimensions (or modes)—and this decomposition is essentially unique under mild conditions. The proposed method employs a coupled CPD model to tackle the HSI-MSI fusion task. We show that the model guarantees the identifiability of the SRI under realistic conditions, leveraging uniqueness of the (coupled) CPD model. Our proposed method is the first identifiability-guaranteed HSR approach. Note that identifiability-guaranteed models and algorithms are not only of theoretical interest—they usually offer more favorable empirical results, e.g. exhibiting enhanced performance, as we will see. Furthermore, the proposed approach can work under scenarios where the spatial degradation operator is unknown. Unlike some existing methods which attempt to estimate the spatial degradation operator [16], [35], our method works under the case where the spatial degradation operator is not known at all—without losing identifiability of the SRI. Numerical experiments using synthetic and semi-real data

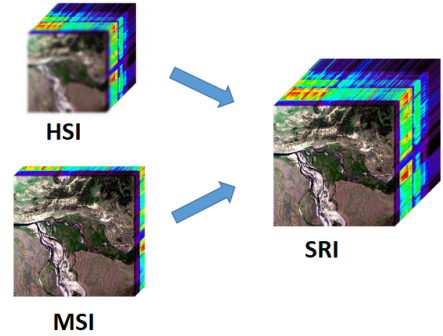

Fig. 1: Illustration of the hyperspectral super-resolution task.

show that the proposed approach is very promising for the hyperspectral super-resolution task.

A preliminary version of part of this work has been accepted in ICASSP 2018. The ICASSP version [36] presents the basic approach and part of the simulations. This journal version additionally includes detailed analysis of the model and the algorithms, proof of the theorems, extensive semi-real data simulations, and more baselines for comprehensive comparison.

## II. PROBLEM STATEMENT AND BACKGROUND

Consider an HSI cube $\underline{\mathbf{Y}}_H \in \mathbb{R}^{I_H \times J_H \times K_H}$, where $I_H$ and $J_H$ denote the spatial dimensions and $K_H$ denotes the number of spectral bands. Similarly, let $\underline{\mathbf{Y}}_M \in \mathbb{R}^{I_M \times J_M \times K_M}$ denote an MSI cube, where $I_M, J_M$ and $K_M$ are the dimensions of the spatial and spectral domains, respectively. An HSI captures information over a broad range of the electromagnetic spectrum, usually involving hundreds of spectral bands/wavelengths. An MSI usually consists of pixels which are measured at less than 20 wavelengths; i.e., $K_M \ll K_H$ in general. On the other hand, MSIs have a much finer resolution in the spatial domain relative to HSIs—i.e., $I_H J_H \ll I_M J_M$ typically holds.

Hyperspectral super-resolution aims at integrating a pair of co-registered HSI and MSI, which describe the same target (e.g., a region on the ground), in order to form an SRI $\underline{\mathbf{Y}}_S \in \mathbb{R}^{I_M \times J_M \times K_H}$ that has the spatial resolution of the MSI and the spectral resolution of the HSI. The hyperspectral super-resolution task, illustrated in Fig. 1, is very well-motivated since both spectral and spatial information are rich and valuable to analytics and can benefit a number of applications such as image processing, remote sensing, geoscience, and food and medicine security, just to name a few.

### A. Matrix Factorization-based Approaches

The arguably most popular and effective existing HSR approaches are based on low-rank matrix factorization. Specifically, in [16]–[21], [23], the matricized multiband images (i.e., SRI, HSI, MSI) are all modeled as low rank matrices, resulting from the linear mixure model (LMM) of the multiband pixels. To be specific, consider the matricized SRI as:

$$\mathbf{Y}_S = [\underline{\mathbf{Y}}_S(1,1,:),\ldots,\underline{\mathbf{Y}}_S(I_H, J_H,:)]^T \in \mathbb{R}^{I_M J_M \times K_H}, \quad (1)$$

where $\underline{\boldsymbol{Y}}_S(i,j,:) \in \mathbb{R}^{K_H}$ is a vector that is formed by taking the $(i,j)$th spectral pixel of the SRI. Under the LMM, a spectral pixel $\boldsymbol{Y}_S(:,\ell)$ is modeled as a weighted sum of the spectral signatures of several materials (or endmembers) that are present in the image:

$$\boldsymbol{Y}_S \approx \boldsymbol{S}_M \boldsymbol{E}_H^T, \quad (2)$$

where $\boldsymbol{E}_H \in \mathbb{R}^{K_H \times R}$ is the endmember matrix containing the spectral signatures of $R \ll \min\{I_M J_M, K_H\}$ materials in its $R$ columns and $\boldsymbol{S}_M \in \mathbb{R}^{I_H J_H \times R}$ is the abundance matrix.

In order to tackle the HSR problem, existing work usually assumes that there exist two linear operators $\boldsymbol{P}_H \in \mathbb{R}^{I_H J_H \times I_M J_M}$ and $\boldsymbol{P}_M \in \mathbb{R}^{K_M \times K_H}$ such that $\boldsymbol{Y}_H = \boldsymbol{P}_H \boldsymbol{Y}_S$ and $\boldsymbol{Y}_M = \boldsymbol{Y}_S \boldsymbol{P}_M^T$. As a result, the matricized HSI is modeled as $\boldsymbol{Y}_H = (\boldsymbol{P}_H \boldsymbol{S}_M) \boldsymbol{E}_H^T$ and the matricized MSI as $\boldsymbol{Y}_M = \boldsymbol{S}_M (\boldsymbol{P}_M \boldsymbol{E}_H)^T$. Then, if $\boldsymbol{E}_H$ and $\boldsymbol{S}_M$ (or the range spaces of $\boldsymbol{E}$ and $\boldsymbol{S}$) can be estimated via jointly factoring $\boldsymbol{Y}_H$ and $\boldsymbol{Y}_M$ following the described model, the SRI is recovered following equation (2). This is the basic idea behind the low-rank factorization based HSR approaches.

*B. Challenges*

The low rank matrix factorization approaches are effective to a certain extent and considered state of the art. However, three key theoretical and practical challenges remain.

First, as previously mentioned, multiband images are naturally data cubes that exhibit dependence across all the three dimensions. Using the matricized version of the 3D images is prone to loss of structural information. Some existing works tried to compensate this loss of information via promoting spatial smoothness (e.g., by adding total variation constraints on $\boldsymbol{S}_M$ or $\boldsymbol{S}_H$ [16]). This is a viable solution but several issues remain—e.g., these methods have to introduce a few more tuning parameters that are in general hard to determine. In addition, merely using spatial smoothness still cannot fully exploit the dependence across the two spatial dimensions and the spectral dimension.

The second challenge is that recovering $\boldsymbol{Y}_S$ from the matricized HSI and MSI, i.e., $\boldsymbol{Y}_H$ and $\boldsymbol{Y}_M$, is an ill-posed inverse problem—an infinite number of solutions could exist. Making use of the low-rank modeling could help reduce the difficulty since it reduces the number of unknowns substantially—but there is still a lack of theoretical evidence that this approach can recover $\boldsymbol{Y}_S$ correctly. One possible route for arguing identifiability is to connect the matrix factorization-based approaches to low-rank matrix sensing. However, this would require the degradation operators to be random [37], [38]. In our context, the degradation operators are highly structured, which means that known theory of matrix sensing cannot answer our question. The coupled factorization approaches with a variety of regularizations [16], [17], [19], [20] may help in practice—but currently lack theoretical identifiability guarantees. Note that identifiability is also important from a practical viewpoint, apart form theoretical. In particular, identifiability often serves as a guidance for practitioners to select and design the appropriate solvers and algorithms—which have been proven very useful and powerful in pertinent problems, such as hyperspectral unmixing [27]. Furthermore, it has been often observed that, in a variety of problems (such as matrix and tensor decomposition), identifiability-guaranteed criteria usually entail stable numerical performance [30], [33].

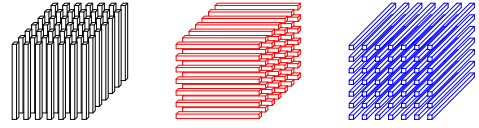

Fig. 2: The columns ($\underline{\boldsymbol{X}}(:,j,:)$), rows ($\underline{\boldsymbol{X}}(i,:,:)$), and fibers ($\underline{\boldsymbol{X}}(:,:,k)$) of a third-order tensor, respectively.

Another major concern is that the matrix-based methods commonly assume that the degradation operators $\boldsymbol{P}_H$ and $\boldsymbol{P}_M$ are accurately known or can be easily estimated, which is hardly the case in practice. The spectral response $\boldsymbol{P}_M$ can be relatively easy to model and estimate by comparing the spectral specifications of the hyperspectral and multispectral sensors. However, modeling the spatial operator can be rather difficult. One commonly used model assumes that the transformation from SRI to HSI is a combination of blurring by a Gaussian kernel and a downsampling process. This is of course a rough approximation and may be far from being accurate. Even if this assumption is approximately true, there is still a number of uncertainties such as the blurring function, the kernel size and the sampling offset. There are approaches in the literature, e.g., [16], [35], that attempt to estimate the degradation operators from data, but, again, these methods have to make a number of model assumptions regarding the degradation process, which can only approximately hold to some extent.

## III. TENSOR ALGEBRA PRELIMINARIES

In this work, we propose a new approach that leverages powerful analytical tools from tensor algebra. Consequently, the proposed fusion method is able to circumvent the aforementioned issues that arise in the matrix factorization based approaches. To facilitate our upcoming discussion, let us briefly introduce some pertinent key concepts of tensor algebra in this section. A third-order tensor $\underline{\boldsymbol{X}} \in \mathbb{R}^{I \times J \times K}$ can be considered as a three-way array indexed by $i, j, k$ with elements $\underline{\boldsymbol{X}}(i,j,k)$. It consists of three modes: columns $\underline{\boldsymbol{X}}(i,:,k)$, rows $\underline{\boldsymbol{X}}(:,j,k)$ and fibers $\underline{\boldsymbol{X}}(i,j,:)$; see Fig. 2. A rank-one tensor $\underline{\boldsymbol{Z}} \in \mathbb{R}^{I \times J \times K}$ is the outer product of three vectors, i.e., $\underline{\boldsymbol{Z}}(i,j,k) = \boldsymbol{a}(i)\boldsymbol{b}(j)\boldsymbol{c}(k)$ for all $i, j, k$, where $\boldsymbol{a} \in \mathbb{R}^I, \boldsymbol{b} \in \mathbb{R}^J$ and $\boldsymbol{c} \in \mathbb{R}^K$. The shorthand notation for the above is $\underline{\boldsymbol{Z}} = \boldsymbol{a} \circ \boldsymbol{b} \circ \boldsymbol{c}$, where $\circ$ denotes the outer product. The *tensor rank* is the smallest number of outer products (rank-one tensors) needed to synthesize $\underline{\boldsymbol{X}}$. Any tensor can be expressed as a sum of outer products, i.e.,

$$\underline{\boldsymbol{X}} = \sum_{f=1}^{F} \boldsymbol{a}_f \circ \boldsymbol{b}_f \circ \boldsymbol{c}_f. \quad (3)$$

If $F$ denotes the minimum number of outer products needed to express $\underline{\boldsymbol{X}}$, then the tensor rank is $F$, i.e., rank($\underline{\boldsymbol{X}}$) = $F$, and the decomposition is known as *canonical polyadic decomposition* (CPD) [30]. Per tensor element, the CPD model can be written as $\underline{\boldsymbol{X}}(i,j,k) = \sum_{f=1}^{F} \boldsymbol{A}(i,f)\boldsymbol{B}(j,f)\boldsymbol{C}(k,f)$, where $\boldsymbol{A} = [\boldsymbol{a}_1, \ldots, \boldsymbol{a}_F] \in \mathbb{R}^{I \times F}$, $\boldsymbol{B} = [\boldsymbol{b}_1, \ldots, \boldsymbol{b}_F] \in$ $\mathbb{R}^{J \times F}$, $\boldsymbol{C} = [\boldsymbol{c}_1, \ldots, \boldsymbol{c}_F] \in \mathbb{R}^{K \times F}$ are called the low-rank latent factors of the third-order tensor. Since a third-order tensor can be fully characterized by its latent factors, we sometimes use the notation $\underline{\boldsymbol{X}} = [\![\boldsymbol{A}, \boldsymbol{B}, \boldsymbol{C}]\!]$ to represent the tensor.

One nice property of tensors is that the CPD model is essentially unique even when $F$ is much larger than $\max\{I, J, K\}$. This is a striking difference between tensors and matrices—the low-rank decomposition of a matrix is, in general, non-unique. The following theorem provides a basic result on the uniqueness of the CPD model:

**Theorem 1** *[39] Let $\underline{\boldsymbol{X}} = [\![\boldsymbol{A}, \boldsymbol{B}, \boldsymbol{C}]\!]$ with $\boldsymbol{A} : I \times F$, $\boldsymbol{B} : J \times F$, and $\boldsymbol{C} : K \times F$. Assume that $\boldsymbol{A}$, $\boldsymbol{B}$ and $\boldsymbol{C}$ are drawn from some joint absolutely continuous distribution. Also assume $I \geq J \geq K$ without loss of generality. If $F \leq 2^{\lfloor \log_2 J \rfloor + \lfloor \log_2 K \rfloor - 2}$, then the decomposition of $\underline{\boldsymbol{X}}$ in terms of $\boldsymbol{A}, \boldsymbol{B},$ and $\boldsymbol{C}$ is essentially unique, almost surely.*

In cases where the tensor rank $F$ is less than or equal to one of the dimensions, the above conditions can be relaxed to the following:

**Theorem 2** *[39] Let $\underline{\boldsymbol{X}} = [\![\boldsymbol{A}, \boldsymbol{B}, \boldsymbol{C}]\!]$ with $\boldsymbol{A} : I \times F$, $\boldsymbol{B} : J \times F$, and $\boldsymbol{C} : K \times F$. Assume that $\boldsymbol{A}, \boldsymbol{B}$ and $\boldsymbol{C}$ are drawn from some joint absolutely continuous distribution. Also assume $I \geq J \geq K$ without loss of generality and $F \leq I$. If $F \leq \min(I, (J-1)(K-1))$, then the decomposition of $\underline{\boldsymbol{X}}$ in terms of $\boldsymbol{A}, \boldsymbol{B},$ and $\boldsymbol{C}$ is essentially unique, almost surely.*

Here, essential uniqueness means that if $\tilde{\boldsymbol{A}}, \tilde{\boldsymbol{B}}, \tilde{\boldsymbol{C}}$ also satisfy $\underline{\boldsymbol{X}} = [\![\tilde{\boldsymbol{A}}, \tilde{\boldsymbol{B}}, \tilde{\boldsymbol{C}}]\!]$, we can only have $\boldsymbol{A} = \tilde{\boldsymbol{A}} \boldsymbol{\Pi} \boldsymbol{\Lambda}_1$, $\boldsymbol{B} = \tilde{\boldsymbol{B}} \boldsymbol{\Pi} \boldsymbol{\Lambda}_2$, and $\boldsymbol{C} = \tilde{\boldsymbol{C}} \boldsymbol{\Pi} \boldsymbol{\Lambda}_3$, where $\boldsymbol{\Pi}$ is a permutation matrix and $\boldsymbol{\Lambda}_i$ is a full rank diagonal matrix such that $\boldsymbol{\Lambda}_1 \boldsymbol{\Lambda}_2 \boldsymbol{\Lambda}_3 = \boldsymbol{I}$. The CPD uniqueness condition is rather mild: Consider, for example, a $80 \times 80 \times 80$ tensor. Following Theorem 1, it admits an essentially unique CPD representation if $F \leq 1024$. This is far more relaxed compared to uniqueness conditions for matrix factorization, where the rank has to be lower than the matrix dimensions and nonnegativity, sparsity, and geometric conditions are required [31]–[34].

One useful operation in tensor algebra is *matricization*. There are different ways to matricize a third-order tensor with a size of $I \times J \times K$. For example, the mode-3 matricization (unfolding) is as follows:

$$\boldsymbol{X}^{(3)} := [\text{vec}(\underline{\boldsymbol{X}}(:,:,1)), \text{vec}(\underline{\boldsymbol{X}}(:,:,2)), \ldots, \text{vec}(\underline{\boldsymbol{X}}(:,:,K))] \tag{4}$$

where 'vec$(\cdot)$' is the vectorization operator. One can see that $\boldsymbol{X}^{(3)} \in \mathbb{R}^{IJ \times K}$. In this case, the matricized tensor $\boldsymbol{X}^{(3)}$ takes the following form:

$$\boldsymbol{X}^{(3)} = (\boldsymbol{B} \odot \boldsymbol{A}) \boldsymbol{C}^T \quad \in \mathbb{R}^{IJ \times K},$$

where $\odot$ denotes the Kronecker product. The operation in (4) is in fact one of the three commonly used ways to matricize (or unfold) a third-order tensor, and is exactly the same operation as in (1)—where we matricized the images cubes to matrices by re-arranging the pixels. The superscript '(3)' is used to denote that the matricization is operated over the third mode; i.e. the fibers of tensor $\underline{\boldsymbol{X}}$ are the columns of matrix $\boldsymbol{X}^{(3)}$. Matricization over the first and the second modes of are similar, which are shown in the following:

$$\boldsymbol{X}^{(1)} := [\text{vec}(\underline{\boldsymbol{X}}(1,:,:)), \text{vec}(\underline{\boldsymbol{X}}(2,:,:)), \ldots, \text{vec}(\underline{\boldsymbol{X}}(I,:,:))]$$
$$\boldsymbol{X}^{(1)} = (\boldsymbol{C} \odot \boldsymbol{B}) \boldsymbol{A}^T \in \mathbb{R}^{KJ \times I}, \tag{5}$$

and

$$\boldsymbol{X}^{(2)} := [\text{vec}(\underline{\boldsymbol{X}}(:,1,:)), \text{vec}(\underline{\boldsymbol{X}}(:,2,:)), \ldots, \text{vec}(\underline{\boldsymbol{X}}(:,J,:))]$$
$$\boldsymbol{X}^{(2)} = (\boldsymbol{C} \odot \boldsymbol{A}) \boldsymbol{B}^T \in \mathbb{R}^{KI \times J}. \tag{6}$$

Another important operation in tensor analytics is the *mode product*. The mode product operator multiplies a matrix to a tensor in one mode. Recall that a third-order tensor $\underline{\boldsymbol{X}} \in \mathbb{R}^{I \times J \times K}$ has three modes (i.e., rows, columns, fibers—cf. Fig. 2) and therefore is associated three different mode products. Applying mode-1, mode-2, and mode-3 products to a third-order tensor sequentially is represented using the following notation:

$$\tilde{\underline{\boldsymbol{X}}} = \underline{\boldsymbol{X}} \times_1 \boldsymbol{P}_1 \times_2 \boldsymbol{P}_2 \times_3 \boldsymbol{P}_3 \tag{7}$$

where "$\times_1$" denotes the operation that multiplies each column of $\underline{\boldsymbol{X}}$ with $\boldsymbol{P}_1$, "$\times_2$" denotes multiplying each row of $\underline{\boldsymbol{X}}$ with $\boldsymbol{P}_2$, and "$\times_3$" denotes multiplying each fiber of $\underline{\boldsymbol{X}}$ with $\boldsymbol{P}_3$. A very important property of mode product is that the result of (7) is a third-order tensor that can be represented as:

$$\tilde{\underline{\boldsymbol{X}}} = [\![\boldsymbol{P}_1 \boldsymbol{A}, \boldsymbol{P}_2 \boldsymbol{B}, \boldsymbol{P}_3 \boldsymbol{C}]\!],$$

which is polyadic decomposition of $\tilde{\underline{\boldsymbol{X}}}$ with $F$ rank-one tensors, and is an essentially unique decomposition under some conditions—which is a very useful insight, as we will see.

## IV. Degradation as Mode Product

Beginning from this section, we propose a tensor-based approach to handle the HSR problem, which admits an array of good features that the matrix-based methods do not have. In this section, we first reveal a nice connection between tensor mode products and the SRI-HSI/MSI degradation models. Building upon this connection, we will introduce coupled tensor factorization formulations and offer identifiability analyses in the next section.

Let $\underline{\boldsymbol{Y}}_S \in \mathbb{R}^{I_M \times J_M \times K_H}$ be the target SRI we want to estimate. The tensor $\underline{\boldsymbol{Y}}_S$ admits a CPD with rank $F$, i.e.,

$$\underline{\boldsymbol{Y}}_S = [\![\boldsymbol{A}, \boldsymbol{B}, \boldsymbol{C}]\!] \tag{8}$$
$$\boldsymbol{A} : I_M \times F, \ \boldsymbol{B} : J_M \times F, \ \boldsymbol{C} : K_H \times F$$

Also let $\underline{\boldsymbol{Y}}_H \in \mathbb{R}^{I_H \times J_H \times K_H}$ denote the corresponding HSI and $\underline{\boldsymbol{Y}}_M \in \mathbb{R}^{I_M \times J_M \times K_M}$ the MSI, respectively. Assume that there exist $\boldsymbol{P}_1$ and $\boldsymbol{P}_2$ such that the spatial degradation from the SRI to the HSI can be modeled as

$$\underline{\boldsymbol{Y}}_H(:,:,k) = \boldsymbol{P}_1 \underline{\boldsymbol{Y}}_S(:,:,k) \boldsymbol{P}_2^T, \ k = 1, \ldots, K_H, \tag{9}$$

where $\boldsymbol{P}_1 \in \mathbb{R}^{I_H \times I_M}$ and $\boldsymbol{P}_2 \in \mathbb{R}^{J_H \times J_M}$. The degradation model in Eq. (9) is intuitive: Blurring can be modeled as linear mixing of neighboring pixels under a certain kernel in both the column and row dimensions, and downsampling can be viewed as linear compression—and the two procedures can be well modeled using a 'fat' matrix $\boldsymbol{P}_1$ and a 'tall' matrix $\boldsymbol{P}_2$ with appropriate kernels 'embedded' in the matrix elements. In fact, the model in (9) summarizes some popularly used blurring and downsampling models of the spatial degradation process. For example, in Appendix C, we show that the 2-D Gaussian blurring plus downsampling model, which is widely adopted in the HSR literature [16]–[21], [23], can be re-expressed as (9).

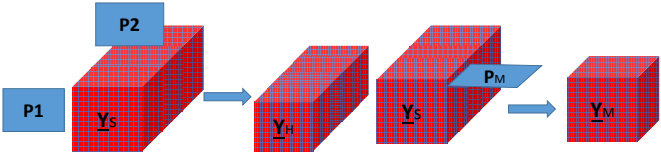

Fig. 3: Illustration of degradation from the super-resolution image to the HSI and MSI, respectively.

Under (9), it is straightforward to observe that the model can be written as $\underline{\boldsymbol{Y}}_H = \underline{\boldsymbol{Y}}_S \times_1 \boldsymbol{P}_1 \times_2 \boldsymbol{P}_2$ (and thus $\boldsymbol{P}_H = \boldsymbol{P}_2 \otimes \boldsymbol{P}_1$ in the matricized form). Consequently, $\underline{\boldsymbol{Y}}_H$ can be represented in the following form:

$$\underline{\boldsymbol{Y}}_H = [\![\tilde{\boldsymbol{A}}, \tilde{\boldsymbol{B}}, \boldsymbol{C}]\!] \tag{10}$$
$$\tilde{\boldsymbol{A}} = \boldsymbol{P}_1 \boldsymbol{A} : I_H \times F, \ \tilde{\boldsymbol{B}} = \boldsymbol{P}_2 \boldsymbol{B} : J_H \times F, \ \boldsymbol{C} : K_H \times F$$

In the matricized form, we have $\boldsymbol{Y}_H^{(3)} = \left(\tilde{\boldsymbol{B}} \odot \tilde{\boldsymbol{A}}\right) \boldsymbol{C}^T$.

The spectral degradation from the SRI to the MSI can be modeled as

$$\underline{\boldsymbol{Y}}_M(i,j,:) = \boldsymbol{P}_M \underline{\boldsymbol{Y}}_S(i,j,:) \ \forall i,j \tag{11}$$

where $\underline{\boldsymbol{Y}}_S(i,j,:) \in \mathbb{R}^K$ represents a fiber of the SRI and $\underline{\boldsymbol{Y}}_M(i,j,:) \in \mathbb{R}^K$ a fiber of the MSI, respectively. Matrix $\boldsymbol{P}_M \in \mathbb{R}^{K_M \times K_H}$ is usually modeled as a band-selection and averaging matrix. Eq. (11) is nothing but a mode-3 product operation, i.e., $\underline{\boldsymbol{Y}}_M = \underline{\boldsymbol{Y}}_S \times_3 \boldsymbol{P}_M$. Hence, $\underline{\boldsymbol{Y}}_M$ can be written as the following:

$$\underline{\boldsymbol{Y}}_M = [\![\boldsymbol{A}, \boldsymbol{B}, \tilde{\boldsymbol{C}}]\!] \tag{12}$$
$$\boldsymbol{A} : I_M \times F, \ \boldsymbol{B} : J_M \times F, \ \tilde{\boldsymbol{C}} = \boldsymbol{P}_M \boldsymbol{C} : K_M \times F$$

It is also readily seen that $\boldsymbol{Y}_M^{(3)} = (\boldsymbol{B} \odot \boldsymbol{A})(\boldsymbol{P}_M \boldsymbol{C})^T$.

The discussed connection between HSR degradation and tensor mode products is visualized in Fig. 3. In retrospect, this connection is not hard to see for someone who is versed in tensor algebra. However, the implication is very interesting and significant: If the "compressed" HSI and MSI tensors admit unique CPD models, then the SRI can be recovered. Intuitively, if one can identify the latent factors of $\underline{\boldsymbol{Y}}_M$ and $\underline{\boldsymbol{Y}}_H$ via CPD, respectively, then the SRI can be reconstructed using $\underline{\boldsymbol{Y}}_S = [\![\boldsymbol{A}, \boldsymbol{B}, \boldsymbol{C}]\!]$. This, of course, is only intuitive and needs to be refined in a number of aspects, but it reveals the major insight that leads to provable identifiability results for the HSR problem—as we will see in the next section.

## V. COUPLED TENSOR FACTORIZATION FOR SUPER-RESOLUTION

Following the insights revealed in the previous section, we develop algorithms to handle the HSR problem in this section. We consider two cases: First, when the degradation operators are known, which follows the standard setups as the majority of matrix-based HSR works e.g., [17]–[20], [24]. Second, when the spatial degradation operator is unknown, which is more realistic yet much more challenging. For both cases, we propose tensor-based algorithms and discuss the respective theoretical guarantees.

### A. When $\boldsymbol{P}_H$ and $\boldsymbol{P}_M$ are known

Let us first consider the case where $\boldsymbol{P}_H$ and $\boldsymbol{P}_M$ are known. Recall that $\underline{\boldsymbol{Y}}_H = [\![\boldsymbol{P}_1 \boldsymbol{A}, \boldsymbol{P}_2 \boldsymbol{B}, \boldsymbol{C}]\!]$ and $\underline{\boldsymbol{Y}}_M = [\![\boldsymbol{A}, \boldsymbol{B}, \boldsymbol{P}_M \boldsymbol{C}]\!]$, where $\tilde{\boldsymbol{A}} = \boldsymbol{P}_1 \boldsymbol{A}$, $\tilde{\boldsymbol{B}} = \boldsymbol{P}_2 \boldsymbol{B}$, and $\tilde{\boldsymbol{C}} = \boldsymbol{P}_M \boldsymbol{C}$. We wish to identify $\boldsymbol{A}$, $\boldsymbol{B}$ and $\boldsymbol{C}$ from the HSI and MSI so that we can reconstruct the SRI. To this end, we propose to employ the following formulation:

$$\underset{\boldsymbol{A},\boldsymbol{B},\boldsymbol{C}}{\text{minimize}} \ \|\underline{\boldsymbol{Y}}_H - [\![\boldsymbol{P}_1 \boldsymbol{A}, \boldsymbol{P}_2 \boldsymbol{B}, \boldsymbol{C}]\!]\|_F^2 \\ + \lambda \|\underline{\boldsymbol{Y}}_M - [\![\boldsymbol{A}, \boldsymbol{B}, \boldsymbol{P}_M \boldsymbol{C}]\!]\|_F^2. \tag{13}$$

In other words, we employ the above formulation to jointly decompose the HSI and MSI tensors to estimate $\boldsymbol{A}$, $\boldsymbol{B}$ and $\boldsymbol{C}$, where $\lambda > 0$ is a pre-selected parameter that weights the importance of each image in estimating $\boldsymbol{A}, \boldsymbol{B}$ and $\boldsymbol{C}$. After obtaining the estimates of $\boldsymbol{A}, \boldsymbol{B}$ and $\boldsymbol{C}$, the super-resolution tensor reconstruction is performed by

$$\hat{\underline{\boldsymbol{Y}}}_S(i,j,k) = \sum_{f=1}^F \hat{\boldsymbol{A}}(i,f)\hat{\boldsymbol{B}}(j,f)\hat{\boldsymbol{C}}(k,f). \tag{14}$$

The problem in (13) is non-convex and also NP-hard in general. To tackle it, the *alternating optimization* (AO) framework is employed. Specifically one factor is updated at a time while keeping the rest fixed. Making use of the matricized forms of the HSI and MSI tensors, every step boils down to solving a Sylvester's equation—which is easy to handle. The proposed *Super-resolution TEnsor-REcOnstruction* (STEREO for short) is summarized in Algorithm 1.

---
**Algorithm 1:** STEREO

**Initialization**: $\lambda$, $F$, $\boldsymbol{A}$, $\boldsymbol{B}$, $\boldsymbol{C}$
**repeat**
  $\boldsymbol{A} \leftarrow \arg\min_{\boldsymbol{A}} \|\boldsymbol{Y}_H^{(1)} - (\boldsymbol{C} \odot \boldsymbol{P}_2 \boldsymbol{B})\boldsymbol{A}^T \boldsymbol{P}_1^T\|_F^2 + \lambda \|\boldsymbol{Y}_M^{(1)} - (\boldsymbol{P}_M \boldsymbol{C} \odot \boldsymbol{B})\boldsymbol{A}^T\|_F^2$;
  $\boldsymbol{B} \leftarrow \arg\min_{\boldsymbol{B}} \|\boldsymbol{Y}_H^{(2)} - (\boldsymbol{C} \odot \boldsymbol{P}_1 \boldsymbol{A})\boldsymbol{B}^T \boldsymbol{P}_2^T\|_F^2 + \lambda \|\boldsymbol{Y}_M^{(2)} - (\boldsymbol{P}_M \boldsymbol{C} \odot \boldsymbol{A})\boldsymbol{B}^T\|_F^2$;
  $\boldsymbol{C} \leftarrow \arg\min_{\boldsymbol{C}} \|\boldsymbol{Y}_H^{(3)} - (\boldsymbol{P}_2 \boldsymbol{B} \odot \boldsymbol{P}_1 \boldsymbol{A})\boldsymbol{C}^T\|_F^2 + \lambda \|\boldsymbol{Y}_M^{(3)} - (\boldsymbol{B} \odot \boldsymbol{A})\boldsymbol{C}^T \boldsymbol{P}_M^T\|_F^2$;
**until** Some stopping criterion is met
Reconstruct $\underline{\boldsymbol{Y}}_S$ using (14).

---

### B. When $\boldsymbol{P}_H$ is unknown

We also consider the case where the spatial degradation operator $\boldsymbol{P}_H = \boldsymbol{P}_2 \otimes \boldsymbol{P}_1$ is completely unknown. As previously explained, considering this scenario is very well-motivated: Although $\boldsymbol{P}_M$ is relatively easy to model since it is well recognized as a uniform spectral response function[1][40], the spatial degradation operator is quite hard to accurately model and estimate. Even when the operation is known as a combination of blurring and downsampling, the hyperparameters such as the blurring kernel type, the kernel size, and the downsampling offset are hardly known in practice. To circumvent this, we propose to employ the following estimator for $A, B, C$:

$$\underset{A,B,\tilde{A},\tilde{B},C}{\text{minimize}} \ \left\| \underline{Y}_H - [\![\tilde{A}, \tilde{B}, C]\!] \right\|_F^2 + \lambda \left\| \underline{Y}_M - [\![A, B, P_M C]\!] \right\|_F^2. \quad (15)$$

---
[1] A reasonable estimate of $\boldsymbol{P}_M$ can usually be obtained after comparing the hyperspectral and multispectral specifications, i.e. the employed wavelengths of the HSI and MSI cameras.

Problem (15) is harder than Problem (13) since it has more unknowns to estimate (as will be reflected in the theoretical analysis in the next subsection). Nevertheless, this problem can still be tackled using AO, just as we handled Problem (13). The `Blind STEREO` algorithm that handles problem (15) is described in Algorithm 2. We use 'Blind' in the algorithm's name to distinguish it with `STEREO`, since Algorithm 2 is *spatially blind*—i.e., it does not need any prior knowledge on the spatial degradation operator.

---

**Algorithm 2:** `Blind STEREO`

**Initialization**: $\lambda$, $F$, $A$, $B$, $\tilde{A}$, $\tilde{B}$
**repeat**
$\quad C \leftarrow \arg\min_C \|Y_H^{(3)} - (\tilde{B} \odot \tilde{A}) C^T\|_F^2 +$
$\qquad\qquad \lambda \|Y_M^{(3)} - (B \odot A) C^T P_3^T\|_F^2;$
$\quad \tilde{A} \leftarrow \arg\min_{\tilde{A}} \|Y_H^{(1)} - (C \odot \tilde{B}) \tilde{A}^T\|_F^2;$
$\quad \tilde{B} \leftarrow \arg\min_{\tilde{B}} \|Y_H^{(2)} - (C \odot \tilde{A}) \tilde{B}^T\|_F^2;$
$\quad C \leftarrow \arg\min_C \|Y_H^{(3)} - (\tilde{B} \odot \tilde{A}) C^T\|_F^2 +$
$\qquad\qquad \lambda \|Y_M^{(3)} - (B \odot A) C^T P_3^T\|_F^2;$
$\quad A \leftarrow \arg\min_A \|Y_M^{(1)} - (P_3 C \odot B) A^T\|_F^2;$
$\quad B \leftarrow \arg\min_B \|Y_M^{(2)} - (P_3 C \odot A) B^T\|_F^2;$
**until** Some stopping criterion is met
Reconstruct $\underline{Y}_S$ using (14).

---

### C. Identifiability Analysis

In this section, we present the identifiability analysis of the proposed approaches. Unlike the matrix factorization-based approaches that mostly have no identifiability characterization of the methods, we show that the proposed estimators can guarantee identifiability of the super-resolution tensor under realistic conditions.

To proceed, let us first consider the following important lemma:

**Lemma 1** *Let $\tilde{Z} = QZ$, where the elements of $Z$ are drawn from an absolutely continuous joint distribution with respect to the Lebesgue measure in $\mathbb{R}^{IF}$ and $Q \in \mathbb{R}^{I' \times I}$ is deterministic with full row rank. Then the joint distribution of the elements in $\tilde{Z}$ is absolutely continuous with respect to the Lebesgue measure in $\mathbb{R}^{I'F}$*

*Proof:* Define $\tilde{z} := \text{vec}(\tilde{Z})$ and $z := \text{vec}(Z)$. Then, we have

$$\tilde{z} = \text{vec}(QZ) = \text{vec}(QZI) = (I \otimes Q) z.$$

Now, define $P = I \otimes Q \in \mathbb{R}^{I'F \times IF}$, which is a 'fat' matrix since $I'F \leq IF$. By properties of the Kronecker product, we have

$$\text{rank}(P) = \text{rank}(Q)\text{rank}(I) = I'F.$$

Furthermore, let $P = U \Sigma V^T$ denote the full-size singular value decomposition (SVD) of $P$, where $U \in \mathbb{R}^{I'F \times I'F}$, $V \in \mathbb{R}^{IF \times IF}$ are orthonormal matrices and $\Sigma \in \mathbb{R}^{I'F \times IF}$ consists of a diagonal submatrix as its first $I'F$ columns (which holds the singular values as the diagonal elements) and an all-zero submatrix, i.e.,

$$\Sigma = [\text{Diag}(\sigma_1, \ldots, \sigma_{I'F}), \mathbf{0}] \in \mathbb{R}^{I'F \times IF}.$$

Consider $z_V = V^T z$ and let $f_Z(z)$ denote the joint probability density function (PDF) of $z$ with respect to the Lebesgue measure in $\mathbb{R}^{IF}$. The random vector $z_V$ is absolutely continuous with respect to the Lebesgue measure in $\mathbb{R}^{IF}$, since the Lebesgue measure is invariant under unitary transformations [41] and the PDF of $z_V$ takes the following form [42]:

$$f_{Z_V}(z_V) = f_Z(V z).$$

Now, consider $z_\Sigma = \Sigma z_V$. This matrix-vector product selects and positively weights the first $I'F$ random variables in $z_V$, i.e.,

$$z_\Sigma = \text{Diag}(\sigma) \tilde{z}_V, \ \tilde{z}_V = z_V(1 : I'F) \in \mathbb{R}^{I'F},$$

where $\sigma = [\sigma_1, \ldots, \sigma_{I'F}]^T$. The above product does not hurt the continuity of the joint distribution of the random variables in $\tilde{z}_V$ (since the joint PDF of $\tilde{z}_V$ can be obtained via marginalizing the joint PDF of $z_V$), and thus $z_\Sigma$ is absolutely continuous with respect to the Lebesgue measure in $\mathbb{R}^{I'F}$. Finally, consider $\tilde{z} = U z_\Sigma$. Again, $\tilde{z}$ is absolutely continuous with respect to the Lebesgue measure in $\mathbb{R}^{I'F}$, since $U$ is a unitary transformation and the PDF is

$$f_{\tilde{Z}}(\tilde{z}) = f_{Z_\Sigma}(U^T z_\Sigma). \qquad \blacksquare$$

With Lemma 1 in our hands, we can show identifiability of the formulations in (13)-(15). To see this, let us first consider the case where the spatial and spectral degradation operators are known. Regarding the identifiability of the SRI cube, let us make some model assumptions to simplify the analysis. We first assume that $I_M \geq J_M \geq K_M$ since $K_M$ is usually quite small (i.e., usually being a single digit) and $I_H \geq J_H$. The number of hyperspectral bands, i.e., $K_H$, could be larger than $I_H$ and $J_H$, depending on how large is the spatial area that we are interested in. Bearing these in mind, we have the following theorem:

**Theorem 3** *Assume that $\underline{Y}_S = [\![A, B, C]\!]$, $\underline{Y}_H = [\![P_1 A, P_2 B, C]\!]$ and $\underline{Y}_M = [\![A, B, P_M C]\!]$. In addition, assume that $I_M \geq J_M \geq K_M$, that $A$, $B$ and $C$ are drawn from some absolutely continuous distribution with respect to the Lebesgue measure in $\mathbb{R}^{(I_M + J_M + K_H)F}$, that $P_1$, $P_2$ and $P_M$ have full rank, and that $(A^\star, B^\star, C^\star)$ is an optimal solution to Problem (13) for some $\lambda > 0$. Then,*

$$\hat{\underline{Y}}_S(i, j, k) = \sum_{f=1}^F A^\star(i, f) B^\star(j, f) C^\star(k, f)$$

*recovers the ground-truth $\underline{\boldsymbol{Y}}_S$ almost surely if*

$$F \leq \min\{2^{\lfloor \log_2(K_M J_M) \rfloor - 2}, I_H J_H\}.$$

The proof is relegated to Appendix A. We should mention that the above bound is proven by judiciously combining Theorem 1, Lemma 1, and the problem structure—and the bound can be improved if $I_H \geq F$ holds. Specifically, we have:

**Corollary 1** *Under the same assumptions as in Theorem 3, if $I_M \geq F$, we have that $\hat{\underline{\boldsymbol{Y}}}_S(i,j,k) = \sum_{f=1}^{F} \boldsymbol{A}^\star(i,f) \boldsymbol{B}^\star(j,f) \boldsymbol{C}^\star(k,f)$ recovers the ground-truth $\underline{\boldsymbol{Y}}_S$ almost surely if $F \leq \min\{(J_M - 1)(K_M - 1), I_H J_H\}$.*

The proof of Corollary 1 is almost the same as that of Theorem 3. The only difference is that Theorem 2 (instead of Theorem 1) is invoked. The proof is omitted due to space limitation.

For the case where $\boldsymbol{P}_1$ and $\boldsymbol{P}_2$ are unknown, we have the following theorem:

**Theorem 4** *Assume the same generative model as in Theorem 3, that $I_M \geq J_M \geq K_M$ and $I_H \geq J_H$, that $I_M J_M \geq I_H J_H$ and $K_M \leq K_H$, and that $(\tilde{\boldsymbol{A}}^\star, \tilde{\boldsymbol{B}}^\star, \boldsymbol{A}^\star, \boldsymbol{B}^\star, \boldsymbol{C}^\star)$ is an optimal solution to Problem (15), for some $\lambda > 0$. Then, $\hat{\underline{\boldsymbol{Y}}}_S(i,j,k) = \sum_{f=1}^{F} \boldsymbol{A}^\star(i,f) \boldsymbol{B}^\star(j,f) \boldsymbol{C}^\star(k,f)$ recovers the ground-truth $\underline{\boldsymbol{Y}}_S$ almost surely,*

1) *if $F \leq \min\{2^{\lfloor \gamma_1 \rfloor - 2}, 2^{\lfloor \gamma_2 \rfloor - 2}\}$, where $\gamma_1 = \log_2(J_M K_M)$ and $\gamma_2 = \log_2(J_H K_H)$, when $I_H \geq K_H$; and*
2) *if $F \leq \min\{2^{\lfloor \gamma_1 \rfloor - 2}, 2^{\lfloor \gamma_2 \rfloor - 2}\}$, where $\gamma_1 = \log_2(J_M K_M)$ and $\gamma_2 = \log_2(I_H J_H)$, when $J_H < K_H$.*

Note that if $I_M \geq F$, $2^{\lfloor \gamma_1 \rfloor - 2}$ can be replaced by $(J_M - 1)(K_M - 1)$. Similarly, if $I_H \geq F$, $2^{\lfloor \gamma_2 \rfloor - 2}$ can be replaced by $(J_H - 1)(\min\{I_H, K_H\} - 1)$. The proof of Theorem 4 is relegated to Appendix B. Note that Theorem 3 only requires that the CPD of the MSI tensor is unique, and has more relaxed conditions compared to those in Theorem 4, which needs the CPDs of both the HSI and MSI to be unique. This echoes our comment that Problem (15) is harder than Problem (13), since the former works under the case where one knows less about the model.

To get a concrete sense of the theorems, consider the case where we intend to reconstruct an SRI of size $600 \times 520 \times 180$ from an HSI of size $150 \times 130 \times 180$ and an MSI of size $600 \times 520 \times 8$. By Theorems 3 - 4, the identifiability of the SRI is guaranteed if the CPD rank of the SRI tensor satisfies $F \leq 1024$. This is in general easy to satisfy (approximately) in practice. To verify this, in Table I, we use a CPD model to reconstruct a subimage of the real-world Cuprite hyperspectral image captured by the AVIRIS hyperspectral sensor [43] in Nevada, United States. One can see that the fitting error, defined as $\|\hat{\underline{\boldsymbol{Y}}} - \underline{\boldsymbol{Y}}\|_F / \|\underline{\boldsymbol{Y}}\|_F$ (where $\hat{\underline{\boldsymbol{Y}}}$ and $\underline{\boldsymbol{Y}}$ are the CPD model approximated HSI and the original HSI, respectively), is rather small for all tested ranks (under all these ranks the CPD is unique). Table III shows that using an identifiable CPD model to approximate real-world hyperspectral/multispectral images is very reasonable.

TABLE I: The NMSE of using a CPD model to approximate a subimage of the AVIRIS Cuprite data that is of size $512 \times 614 \times 187$.

| rank | 300 | 400 | 500 | 600 | 700 | 800 |
|---|---|---|---|---|---|---|
| fitting error | 0.019 | 0.016 | 0.0142 | 0.0131 | 0.0123 | 0.0116 |

## VI. SIMULATIONS

In this section, we showcase the effectiveness of the proposed HSR framework using numerical experiments. We generate simulated HSI and SRI images following the Wald's protocol [44]. In Wald's protocol, the SRI-HSI degradation consists of spatial blurring by a convolutional kernel and a downsampling procedure. In order to obtain an MSI from an SRI, the spectral specifications of the multispectral sensor are used, which in our experiments are taken from the LANDSAT [45] or the QuickBird sensor [46]. The LANDSAT sensor produces a 6-band MSI by capturing information in the following spectral bands: Blue (450 - 520 nm), Green (520 - 600 nm), Red (630 - 690 nm), Near-IR (760 - 900 nm), Shortwave-IR1 (1550 - 1750 nm), Shortwave-IR2 (2080 - 2350 nm), whereas the QuickBird sensor produces a 4-band MSI in Blue (430 - 545 nm), Green (466 - 620 nm), Red (590 - 710 nm) and Near-IR (715 - 918 nm). Then, the specifications of the available SRI, which span the spectrum from 400nm to 2500nm in our experiments, are compared with the multispectral sensor bands to form spectral response matrix $\boldsymbol{P}_M$ and thus the tested MSI images. To be more precise, $\boldsymbol{P}_M$ is a selection-averaging matrix which acts on the common wavelengths of the SRI and MSI.

*1) Baselines:* A set of baseline algorithms are employed for comparison, namely, FUSE [18], FUSE-Sparse [19], FUMI [20], HySure [16] and CNMF [17]—which have all demonstrated competitive performance in the literature. All simulations are performed in MATLAB on a Linux server with 3.6GHz cores and 32GB RAM. We propose two CPD based algorithms to cleverly initialize STEREO and Blind STEREO. The idea is to compute the CPD of $\underline{\boldsymbol{Y}}_M$ in order to retrieve $\boldsymbol{A}$, $\boldsymbol{B}$ and then solve a least squares problem to obtain $\boldsymbol{C}$ from $\underline{\boldsymbol{Y}}_H$. This way, an initial guess of the latent factors can be obtained. Consequently, the operational time of the algorithms can be substantially reduced, and an enhanced super-resolution accuracy is empirically observed. Detailed description of the initialization techniques are relegated to Appendix D. The CPD part performed in the initialization procedure is computed using Tensorlab [47] with 25 iterations at maximum. In all the simulations, we fix $\lambda = 1$.

*2) Evaluation:* We largely follow the established conventions in the HSR literature for evaluating the results. Specifically, we adopt several intuitive metrics introduced in [6]. The first metric is *cross correlation (CC)* that is defined as

$$\text{CC} = \frac{1}{K} \sum_{k=1}^{K} \rho(\underline{\boldsymbol{Y}}_S(:,:,k), \hat{\underline{\boldsymbol{Y}}}_S(:,:,k))$$

where $\rho$ is the Pearson correlation coefficient between the estimated and the reference slabs (i.e., $\hat{\underline{\boldsymbol{Y}}}_S(:,:,k)$ and $\underline{\boldsymbol{Y}}_S(:,:,k)$, respectively). CC is a score between 0 and 1, and 1 corresponds to the best estimation result. The second metric is called the *spectral angle mapper (SAM)*, whose definition is as follows:

$$\text{SAM} = \frac{1}{IJ} \sum_{n=1}^{IJ} \arccos\left( \frac{\boldsymbol{Y}_S^{(3)}(n,:)\hat{\boldsymbol{Y}}_S^{(3)}(n,:)^T}{\|\boldsymbol{Y}_S^{(3)}(n,:)\|_2 \|\hat{\boldsymbol{Y}}_S^{(3)}(n,:)\|_2} \right)$$

where $\boldsymbol{Y}_S^{(3)}(n,:)$ and $\hat{\boldsymbol{Y}}_S^{(3)}(n,:)$ represent the corresponding fibers of the ground-truth and the estimated super-resolution tensors, respectively. SAM measures the angles between the estimated and the ground-truth fibers of the SRI, and small SAMs correspond to good performance. The *Relative dimensional global error (ERGAS)* [3] is also employed, which is defined as

$$\text{ERGAS} = 100d\sqrt{\frac{1}{IJK}\sum_{k=1}^{K}\frac{\|\hat{\underline{\boldsymbol{Y}}}_S(:,:,k) - \underline{\boldsymbol{Y}}_S(:,:,k)\|_F^2}{\mu_k^2}},$$

where $d = \frac{I_M}{I_H} = \frac{J_M}{J_H}$ and $\mu_k$ is the mean of the elements in $\underline{\boldsymbol{Y}}_S(:,:,k)$—and small ERGAS values are desired. In addition to the above quality measures, we also employ the *reconstruction Signal-to-Noise ratio (R-SNR)* criterion, i.e.,

$$\text{R-SNR} = 10\log_{10}\left(\frac{\sum_{k=1}^{K}\|\underline{\boldsymbol{Y}}_S(:,:,k)\|_F^2}{\sum_{k=1}^{K}\|\hat{\underline{\boldsymbol{Y}}}_S(:,:,k) - \underline{\boldsymbol{Y}}_S(:,:,k)\|_F^2}\right),$$

and high R-SNR values indicate good reconstruction performance.

### A. Semi-Real Data Experiments

In this subsection, we test the proposed methods under the assumption that both $\boldsymbol{P}_M$ and $\boldsymbol{P}_H$ are known. A real hyperspectral image is used to act as the SRI in our simulations. This way, the 'ground-truth' SRI is known so that the performance can be measured. The corresponding HSI and MSI are degraded from this SRI following Wald's protocol [44] as described before. The degradation process from the SRI to the HSI is modeled as a combination of spatial blurring by a $9 \times 9$ Gaussian kernel and downsampling the blurred image by a factor of $d = 4$ along the two spatial directions.

The first experiment is performed using the dataset that is a subscene of SALINAS HSI from the AVIRIS platform. This scene describes a field that consists of 6 different agricultural products. The image is measured at 224 spectral bands. After removing 20 bands corrupted by water absorption we obtain an 'SRI' of $80 \times 84$ pixels with 204 bands, i.e. $\underline{\boldsymbol{Y}}_S \in \mathbb{R}^{80 \times 84 \times 204}$. Then, $\underline{\boldsymbol{Y}}_H \in \mathbb{R}^{20 \times 21 \times 204}$ is produced through the aforementioned spatial degradation, and $\underline{\boldsymbol{Y}}_M \in \mathbb{R}^{80 \times 84 \times 6}$ is produced through LANDSAT spectral degradation. The rank used in the tensor decomposition is $F = 100$. For the matrix factorization methods, the number of endmembers (model rank) is set to be $R = 6$—which is equal to the ground-truth number of materials.

Table II shows the performance of the algorithms. It is clear that STEREO significantly outperforms the benchmarks. Particularly, in terms of R-SNR, STEREO outperforms FUMI, which admits the best R-SNR among the baselines, by 10 dB. Furthermore, the execution time of the proposed algorithms is very low ($\sim$ 1.3 sec.—similar to most of the matrix based methods), which makes the tensor based approach rather appealing. We also visualize one band of the estimated SRI in

TABLE II: SALINAS scene

| Algorithm | R-SNR | CC | SAM | ERGAS | runtime (sec) |
|---|---|---|---|---|---|
| STEREO | **38.62** | **0.9829** | 0.5495 | **1.3844** | 1.3 |
| FUSE | 28.71 | 0.9174 | 0.4234 | 5.7135 | **0.07** |
| FUSE-Sparse | 28.71 | 0.9173 | 0.4234 | 5.7135 | 69.7 |
| FUMI | 29.40 | 0.9126 | 0.7975 | 6.3527 | 1.56 |
| HySure | 26.86 | 0.8981 | 1.5209 | 6.4187 | 1.6 |
| CNMF | 25.48 | 0.9013 | 1.3225 | 6.3787 | 1.7 |

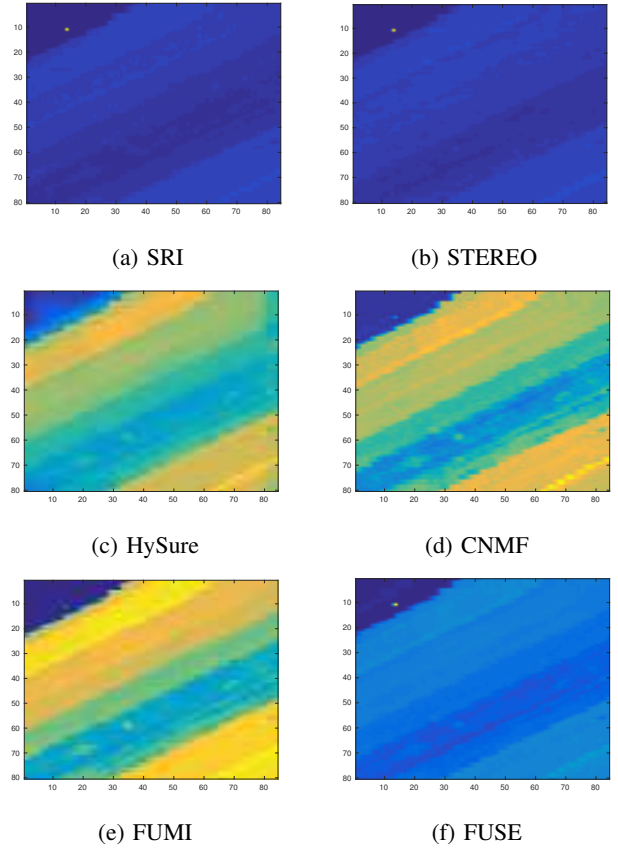

(a) SRI  (b) STEREO
(c) HySure  (d) CNMF
(e) FUMI  (f) FUSE

Fig. 4: SALINAS Reconstruction, 1442nm band

Fig. 4. One can see that the image produced by STEREO is indeed visually significantly closer to the ground-truth SRI.

The second experiment tests the super-resolution methods under scenarios where the degradation models are noisy. The Cuprite HSI downloaded from the AVIRIS platform is used to act as the SRI. The employed subimage has 187 bands (after removing bands corrupted by water absorption) and describes a spatial area containing $512 \times 614$ pixels, i.e., $\underline{\boldsymbol{Y}}_S \in \mathbb{R}^{512 \times 614 \times 187}$. The HSI and MSI are generated as follows:

$$\underline{\boldsymbol{Y}}_H = \underline{\boldsymbol{Y}}_S \times_1 \boldsymbol{P}_1 \times_2 \boldsymbol{P}_2 + \underline{\boldsymbol{N}}_H$$
$$\underline{\boldsymbol{Y}}_M = \underline{\boldsymbol{Y}}_S \times_3 \boldsymbol{P}_M + \underline{\boldsymbol{N}}_M,$$

where $\underline{\boldsymbol{N}}_H$ and $\underline{\boldsymbol{N}}_M$ are additive white Gaussian noise. The degradation operators $\boldsymbol{P}_H = \boldsymbol{P}_2 \otimes \boldsymbol{P}_1$ and $\boldsymbol{P}_M$ are created as before, leading to $\underline{\boldsymbol{Y}}_H \in \mathbb{R}^{128 \times 152 \times 187}$ and $\underline{\boldsymbol{Y}}_M \in \mathbb{R}^{512 \times 614 \times 6}$, respectively. The signal-to-noise ratio (SNR) is

defined as:

$$\text{SNR} = 10 \log_{10} \left( \frac{\sum_{k=1}^{K} \|\underline{\boldsymbol{Y}}(:,:,k)\|_F^2}{\sum_{k=1}^{K} \|\underline{\boldsymbol{N}}(:,:,k)\|_F^2} \right),$$

where $(\underline{\boldsymbol{Y}}, \underline{\boldsymbol{N}})$ stands for either $(\underline{\boldsymbol{Y}}_H, \underline{\boldsymbol{N}}_H)$ or $(\underline{\boldsymbol{Y}}_M, \underline{\boldsymbol{N}}_M)$. The algorithms are examined under different SNRs. In all cases, the SNRs of the HSI and MSI are assumed to be the same. The rank used for tensor decomposition is chosen following Theorems 2-3 and adjusted according to the SNR of each scenario. Precisely, as the SNR varies from 50dB to 20dB the tensor rank changes from $F = 750$ to $F = 100$. The intuition is to use fewer canonical dimensions when the noise level is higher, so that the noise corruption can be better discounted. The rank of the low-rank matrix models is set to be $R = 10$, which is determined by the number of materials in the images. Results are averaged over 10 Monte-Carlo simulations.

Fig. 5 shows the R-SNR performance of the methods under different noise levels. One can see that under high SNR scenarios the proposed STEREO algorithm exhibits the best performance. The matrix-based methods also work very well for the Cuprite data when the noise is almost absent. However, when the SNR drops under 30dB, STEREO vastly outperforms the baselines—which shows the robustness of the proposed method to modeling mismatches. FUSE-Sparse fails to operate due to memory overflow. FUSE seems to be the most vulnerable under noise, and HySure works best among the baselines. The rest of the evaluation metrics are shown in Fig. 6, from which similar a conclusion can be drawn. In terms of the runtime performance, FUSE is the most efficient algorithm, since it only involves very simple procedures. Nevertheless, its accuracy performance is heavily affected by model mismatches in the degradation process, which is undesired in practice. Among the rest, the proposed approach admits the lowest execution time. Note that when the noise level increases, there is a decreasing trend in the runtime of the tensor methods. This happens because the rank reduced when the noise level increased, and lower ranks demand less computational time for the tensor approach.

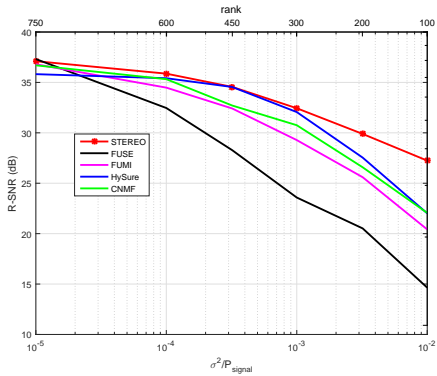

Fig. 5: R-SNR of the algorithms on Cuprite under different noise levels.

Table III shows the performance of the algorithms when SNR=25dB. The tensor rank is set to be $F = 200$ in this case. STEREO produces the best results under all the evaluation metrics. Under this setup, HySure shows the best performance over all the evaluation metrics among the baseline algorithms, but it needs 3 times more runtime compared to that of STEREO. FUSE has the lowest runtime, but the R-SNR is 9dB worse relative to STEREO.

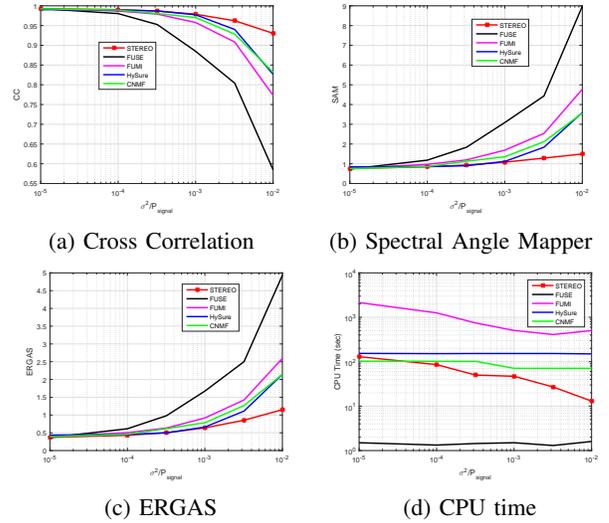

(a) Cross Correlation  (b) Spectral Angle Mapper

(c) ERGAS  (d) CPU time

Fig. 6: Reconstruction metrics for Cuprite

TABLE III: Performance of the algorithms on the Cuprite data. SNR=25dB; "-" means "out of memory".

| Algorithm | R-SNR | CC | SAM | ERGAS | runtime (sec) |
|---|---|---|---|---|---|
| STEREO | **29.89** | **0.96216** | **1.2865** | **0.85327** | 27 |
| FUSE | 20.52 | 0.80408 | 4.4339 | 2.5012 | **1.3** |
| FUSE-Sparse | - | - | - | - | - |
| FUMI | 25.80 | 0.9093 | 2.4404 | 1.3802 | 376.8 |
| HySure | 27.54 | 0.94004 | 1.8377 | 1.11 | 153 |
| CNMF | 26.58 | 0.92826 | 2.1168 | 1.2671 | 71 |

The algorithms are also tested on two more datasets. The first scene, namely, the Indian Pines, was again captured by AVIRIS and contains agriculture, forest and other natural perennial vegetation. The number of ground-truth materials is $R = 16$, and the pixels are measured at 200 bands (after removing water corrupted ones). The SRI in the experiment has $144 \times 144$ pixels, i.e., $\underline{\boldsymbol{Y}}_S \in \mathbb{R}^{144 \times 144 \times 200}$. The HSI and MSI are generated as before, leading to $\underline{\boldsymbol{Y}}_H \in \mathbb{R}^{36 \times 36 \times 200}$ and $\underline{\boldsymbol{Y}}_M \in \mathbb{R}^{144 \times 144 \times 6}$. Table IV shows the performance when the SNR is 25 dB. The tensor rank is $F = 50$. Again, STEREO outperforms the baselines significantly.

TABLE IV: Performance of the algorithms for Indian Pines data. SNR=25dB.

| Algorithm | R-SNR | CC | SAM | ERGAS | runtime (sec) |
|---|---|---|---|---|---|
| STEREO | **25.80** | **0.8077** | **2.5217** | **1.3013** | 1.8 |
| FUSE | 17.29 | 0.5518 | 6.8166 | 4.0767 | **0.1** |
| FUSE-Sparse | 17.29 | 0.5520 | 6.8178 | 4.0783 | 134.5 |
| FUMI | 23.54 | 0.7593 | 3.3931 | 1.8151 | 28.8 |
| HySure | 24.56 | 0.7710 | 2.8371 | 1.5938 | 12.1 |
| CNMF | 23.84 | 0.7321 | 3.0184 | 1.8227 | 4.2 |

The other scene is taken from Pavia University in Italy and was captured by the ROSIS sensor. The SRI, HSI, and MSI are with sizes of $608 \times 336 \times 103$, $152 \times 84 \times 103$ and $608 \times 336 \times 4$, respectively, in which we simulate a QuickBird-generated MSI. Table V shows the performance under SNR=25dB. The tensor rank is $F = 400$, and $R = 9$ is the ground-truth number of materials. One can see that STEREO shows superior performance in this simulation as before.

TABLE V: Performance of the algorithms for Pavia University data. SNR=25dB; "-" means "out of memory".

| Algorithm | R-SNR | CC | SAM | ERGAS | runtime (sec) |
|---|---|---|---|---|---|
| STEREO | **22.50** | **0.9830** | 4.551 | **2.6016** | 26.4 |
| FUSE | 21.09 | 0.9753 | 5.536 | 3.4284 | **0.5** |
| FUSE-Sparse | - | - | - | - | - |
| FUMI | 21.56 | 0.9779 | 5.1151 | 3.0908 | 644.2 |
| HySure | 21.18 | 0.9792 | 4.812 | 2.7934 | 82.5 |
| CNMF | 19.93 | 0.9723 | 5.0183 | 3.3947 | 19.2 |

Table VI shows the R-SNR performance of STEREO under different choices of $F$. One can see that under different SNRs, there is always a wide range of $F$'s (spanning several hundreds of consecutive integers) under which the proposed algorithm works similarly. In other words, the performance of the proposed algorithm is not sensitive to the choice of $F$.

TABLE VI: The obtained R-SNRs (dB) using STEREO under different SNRs and $F$'s.

| SNR / $F$ | 100 | 200 | 300 | 400 | 500 | 600 |
|---|---|---|---|---|---|---|
| 20 dB | 27.25 | 26.00 | 23.41 | 21.44 | 19.61 | 18.36 |
| 25 dB | 28.58 | 29.89 | 29.06 | 27.71 | 26.14 | 24.21 |
| 30 dB | 29.05 | 31.66 | 32.44 | 32.13 | 31.28 | 29.47 |
| 40 dB | 29.20 | 32.32 | 33.90 | 34.47 | 34.50 | 33.74 |
| 50 dB | 29.25 | 32.54 | 34.43 | 35.40 | 35.89 | 35.86 |

### B. Unknown Spatial Degradation Operator

In this subsection, we test our proposed Blind STEREO algorithm under the case where the spatial degradation model is unknown. The SRI used are the Indian Pines and Pavia University images as in the previous section. The HSI $\underline{\mathbf{Y}}_H$ is produced by $\underline{\mathbf{Y}}_S$ after $9 \times 9$ *Gaussian* blurring and downsampling and the MSI $\underline{\mathbf{Y}}_M$ is generated according to LANDSAT and QuickBird specifications, for Indian Pines and Pavia University image respectively.

We fist consider a case where the baseline algorithms falsely assume a $5 \times 5$ Gaussian blurring kernel instead of using the correct $9 \times 9$ kernel. Among the baselines, HySure is able to estimate the degradation operators by assuming knowledge of the kernel size and alignment offset hyperparameters. The SNR of the degradation processes is 25dB. Table VII shows the performance of the algorithms under this scenario using the Indian Pines image. The tensor rank is set to be $F = 50$. One can see that the proposed algorithm yields clearly better reconstruction performance under all the metrics. This shows the advantage of Blind STEREO—since it does not need to assume any prior knowledge on $\mathbf{P}_H$, the considered model mismatches do not affect its performance. Fig. 7 visualizes a band of the reconstructed super-resolution images by the algorithms. One can see that Blind STEREO gives visually more pleasing reconstruction relative to the baselines. The reconstruction performance in the Pavia University image is shown in Table VIII, where the tensor rank is $F = 400$. One can see similar results there.

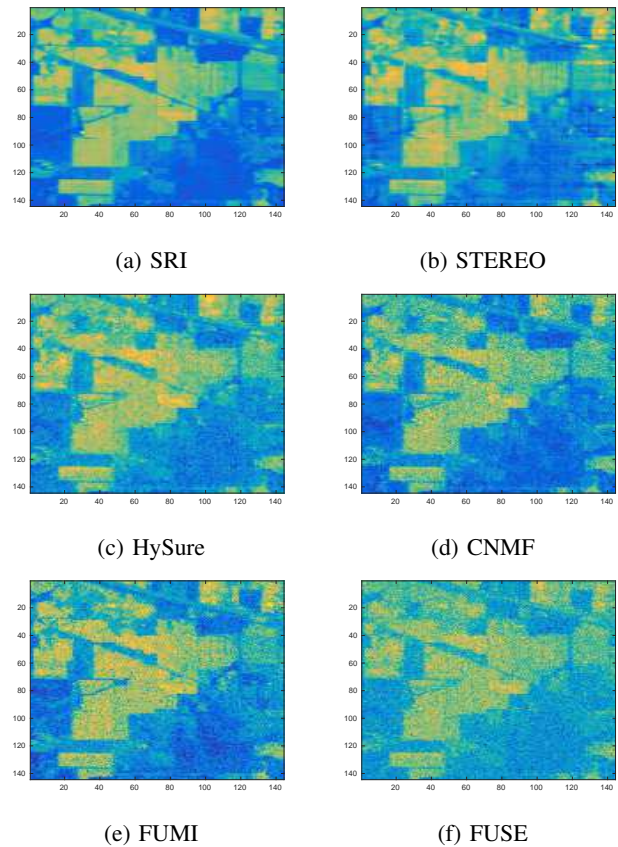

(a) SRI  (b) STEREO

(c) HySure  (d) CNMF

(e) FUMI  (f) FUSE

Fig. 7: Indian Pines Reconstruction, 1422nm band

TABLE VII: Performance of the algorithms on the Indian Pines data under kernel size mismatch

| Algorithm | NMSE | CC | SAM | ERGAS | runtime (sec) |
|---|---|---|---|---|---|
| STEREO | **25.53** | **0.7949** | **2.5831** | **1.3491** | 1 |
| FUSE | 17.51 | 0.5563 | 6.5503 | 3.7642 | **0.13** |
| FUSE-Sparse | 17.75 | 0.5622 | 6.3351 | 3.7574 | 134.4 |
| FUMI | 21.42 | 0.6438 | 4.455 | 2.6748 | 126.1 |
| HySure | 24.64 | 0.7853 | 2.7724 | 1.5255 | 12.4 |
| CNMF | 24.5 | 0.7254 | 3.1102 | 1.8903 | 3.2 |

TABLE VIII: Performance of the algorithms for Pavia University data under kernel size mismatch

| Algorithm | R-SNR | CC | SAM | ERGAS | runtime (sec) |
|---|---|---|---|---|---|
| STEREO | **22.36** | **0.9824** | **4.5997** | **2.6229** | 26 |
| FUSE | 20.83 | 0.97347 | 5.4552 | 3.4906 | **0.5** |
| FUMI | 21.16 | 0.9763 | 5.045 | 3.1508 | 593.3 |
| HySure | 20.68 | 0.9773 | 4.868 | 2.902 | 82.4 |
| CNMF | 19.93 | 0.9727 | 5.0695 | 3.2938 | 19.3 |

We further consider another scenario where the baseline algorithms correctly assume a $9 \times 9$ Gaussian kernel, but the assumed blurring kernel is applied to an area which is misaligned with the ground-truth blurring area by 2 pixels in both of spatial dimensions. Note that such misalignment could easily happen in practice. The results of the second scenario are presented in Tables IX and X. Fig. 8 illustrates the reconstruction performance of Pavia University image at a selected band. Again, one can see that `Blind STEREO` clearly outperforms the benchmarking algorithms.

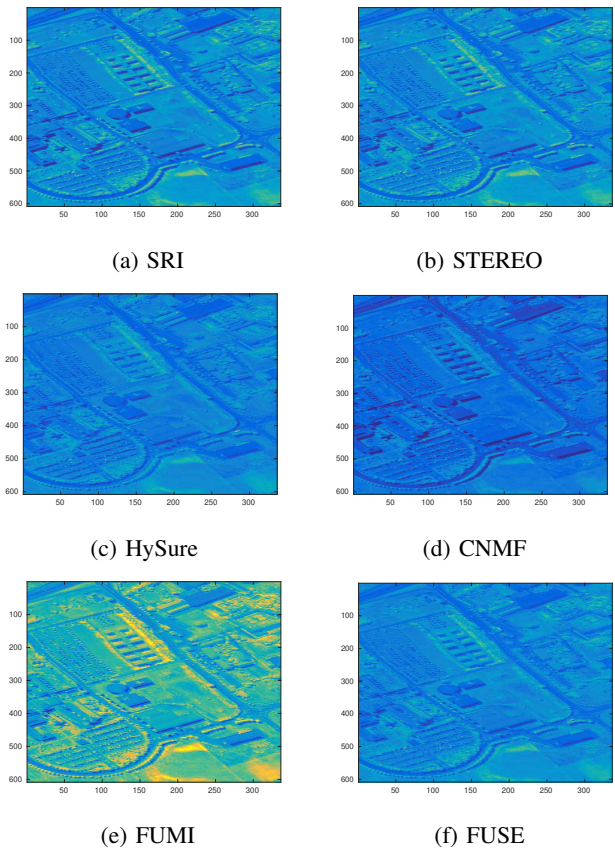

(a) SRI  (b) STEREO
(c) HySure  (d) CNMF
(e) FUMI  (f) FUSE

Fig. 8: Pavia University Reconstruction, 858nm band

TABLE IX: Performance of the algorithms on the Indian Pines data under sampling offset mismatch

| Algorithm | NMSE | CC | SAM | ERGAS | runtime (sec) |
|---|---|---|---|---|---|
| STEREO | **25.81** | **0.8198** | **2.5458** | **1.2788** | 1.5 |
| FUSE | 17.42 | 0.5375 | 6.577 | 3.8526 | **0.14** |
| FUSE-Sparse | 17.72 | 0.5681 | 6.3644 | 3.7683 | 130.5 |
| FUMI | 21.22 | 0.6388 | 4.4524 | 2.5815 | 88.1 |
| HySure | 23.23 | 0.7474 | 3.2332 | 1.7655 | 12.4 |
| CNMF | 24.01 | 0.7339 | 2.9896 | 1.8287 | 5.6 |

TABLE X: Performance of the algorithms for Pavia University data under sampling offset mismatch.

| Algorithm | R-SNR | CC | SAM | ERGAS | runtime (sec) |
|---|---|---|---|---|---|
| STEREO | **22.36** | **0.9824** | **4.5997** | **2.6229** | 26 |
| FUSE | 15.84 | 0.9283 | 7.2734 | 5.2655 | **0.5** |
| FUMI | 16.44 | 0.9392 | 5.8355 | 4.6652 | 287.8 |
| HySure | 17.64 | 0.9571 | 6.4415 | 3.8048 | 82.4 |
| CNMF | 19.93 | 0.9723 | 5.0183 | 3.3947 | 19.2 |

## VII. CONCLUSION

In this work we proposed a novel coupled tensor factorization framework to tackle the hyperspectral super-resolution problem. Compared to the existing matrix-based approaches, the proposed method shows an array of theoretical advantages as well as more promising simulation results. To the best of our knowledge, the developed tensor based HSR framework is the first provably identifiable approach for the very challenging HSR task. One notable feature of the proposed framework is that it can easily accommodate scenarios where the spatial degradation operator is unknown or inaccurately estimated, which is usually the case in practice—without losing identifiability of the SRI. Extensive simulations using a variety of real-world hyperspectral images show that the proposed framework is very promising.

## APPENDIX A
## PROOF OF THEOREM 3

First, we note that for any given $\lambda > 0$, the optimal solution to (13) should make the two terms zero, when the noise is absent. In other words, our problem boils down to considering if the solution to Problem (13) is uniquely determined by $A, B, C$ up to some trivial ambiguities.

Using Lemma 1, it is easily seen that $P_1 A, P_2 B$ and $P_M C$ are drawn from non-singular absolutely continuous distributions. Therefore, Theorems 1 and 2 can be employed to characterize the identifiability of the latent factors of the MSI and HSI tensors. Bearing this in mind, recall that the MSI tensor is derived as $\underline{Y}_M = \underline{Y}_S \times_3 P_M$. We note that under Lemma 1 and the conditions in the statement of Theorem 3, the MSI tensor admits essentially unique latent factors—i.e., if we have $\underline{Y}_M = [\![A_M, B_M, C_M]\!]$, then, the expressions

$$A_M = A\Pi\Lambda_1, \ B_M = B\Pi\Lambda_2, \ C_M = P_M C\Pi\Lambda_3,$$

always hold at an optimal solution of (13) (in the ideal noiseless case), where $\Pi$ is a permutation matrix and $\Lambda_i$ is a full rank diagonal matrix such that $\Lambda_1\Lambda_2\Lambda_3 = I$. In other words, by solving (13) to optimality, $A$ and $B$ can be identified up to column scaling and permutation ambiguities. To establish identifiablity of $C$, let us consider the HSI tensor, i.e.,

$$\underline{Y}_H = \underline{Y}_S \times_1 P_1 \times_2 P_2, \qquad (17)$$

Following the above model, $\underline{Y}_H$ admits a polyadic decomposition (possibly non-unique) $\underline{Y}_H = [\![P_1 A, P_2 B, C_H]\!]$. By matricization, the above can be written as the following:

$$Y_H^{(3)} = (P_2 B \odot P_1 A) C^T, \qquad (18)$$

Plugging in $A_M$ and $B_M$, we have

$$\begin{aligned} Y_H^{(3)} &= (P_2 B_M \odot P_1 A_M) C^T \\ &= (P_2 B \odot P_1 A) \Lambda_1 \Lambda_2 \Pi^2 C^T \\ &= (P_2 B \odot P_1 A) C_H^T \end{aligned} \qquad (19)$$

where $C_H = C\Lambda_3 \Pi$, which is exactly what we wish to identify. The remaining question is whether $C_H$ can be identified from (19)? The answer is affirmative. Indeed, since $P_1 A, P_2 B$ are drawn from absolutely continuous non-singular distributions (cf. Lemma 1), we have $\text{krank}(P_2 B \odot P_1 A) = \min\{I_H J_H, F\}$ almost surely [48]. Then, since $I_H J_H \geq F$, the matrix $P_2 B \odot P_1 A$ has full column rank almost surely and $C_H$ can be uniquely identified from (19).

We should remark that in the proof we did not use identifiablity of the CPD of the HSI tensor. This echoes our comment that even if the CPD of the HSI tensor cannot be uniquely identified, the SRI tensor parametrization can be identified.

## APPENDIX B
## PROOF OF THEOREM 4

The proof is simply by applying Theorem 1 to the HSI and MSI individually. Using the same arguments as in Appendix A the expressions

$$\boldsymbol{A}_M = \boldsymbol{A}\boldsymbol{\Pi}\boldsymbol{\Lambda}_1, \ \boldsymbol{B}_M = \boldsymbol{B}\boldsymbol{\Pi}\boldsymbol{\Lambda}_2, \ \boldsymbol{C}_M = \boldsymbol{P}_M \boldsymbol{C}\boldsymbol{\Pi}\boldsymbol{\Lambda}_3,$$

always hold at an optimal solution of (15). Moreover if $\underline{\boldsymbol{Y}}_H = [\![\boldsymbol{A}_H, \boldsymbol{B}_H, \boldsymbol{C}_H]\!]$ denotes the essentially unique CPD of the HSI tensor, the following expressions

$$\boldsymbol{A}_H = \boldsymbol{P}_1 \boldsymbol{A} \boldsymbol{\Pi}' \boldsymbol{\Lambda}'_1, \ \boldsymbol{B}_H = \boldsymbol{P}_2 \boldsymbol{B} \boldsymbol{\Pi}' \boldsymbol{\Lambda}'_2, \ \boldsymbol{C}_H = \boldsymbol{C} \boldsymbol{\Pi}' \boldsymbol{\Lambda}'_3,$$

also hold at an optimal solution of (15). Note that $\boldsymbol{P}_1$, $\boldsymbol{P}_2$ are unknown. Observe that at an optimal solution of (15) $\boldsymbol{C}_M = \boldsymbol{P}_M \boldsymbol{C}_H$ or equivalently:

$$\boldsymbol{P}_M \boldsymbol{C} \boldsymbol{\Pi} \boldsymbol{\Lambda}_3 = \boldsymbol{P}_M \boldsymbol{C} \boldsymbol{\Pi}' \boldsymbol{\Lambda}'_3.$$

By Lemma 1, the elements of $\boldsymbol{P}_M \boldsymbol{C}$ are drawn from a jointly continuous distribution, thus any square submatrix is full rank almost surely; see [48], for example. It follows that $\boldsymbol{P}_M \boldsymbol{C}$ has Kruskal rank equal to $\min\{K_M, F\} \geq 2$. As a consequence, collinear columns do not exist in $\boldsymbol{P}_M \boldsymbol{C}$ and any column permutation will result in a different matrix. Therefore $\boldsymbol{\Pi} = \boldsymbol{\Pi}'$ and $\boldsymbol{\Lambda}_3 = \boldsymbol{\Lambda}'_3$. Finally, the tensor with CPD factors $\boldsymbol{A}_M$, $\boldsymbol{B}_M, \boldsymbol{C}_H$ is equal to the original tensor $\underline{\boldsymbol{Y}}_S$, i.e. $\hat{\underline{\boldsymbol{Y}}}_S = [\![\boldsymbol{A}_M, \boldsymbol{B}_M, \boldsymbol{C}_H]\!] = [\![\boldsymbol{A}, \boldsymbol{B}, \boldsymbol{C}]\!] = \underline{\boldsymbol{Y}}_S$ almost surely.

## APPENDIX C
## THE SPATIAL DEGRADATION MODEL

The proposed work assumes that the forward spatial degradation from SRI to HSI follows the model in (9), or equivalently that $\boldsymbol{P}_H$ exhibits a Kronecker structure, i.e. $\boldsymbol{P}_H = \boldsymbol{P}_2 \otimes \boldsymbol{P}_1$. Here, we prove that the Kronecker structure assumption on $\boldsymbol{P}_H$ is a generalization of the heavily used 2D Gaussian blurring and downsampling procedure, modeled by $\boldsymbol{P}_H \boldsymbol{Y}_S$ in the matricized form [17]–[20]. To this end, we show that blurring an image by a Gaussian kernel and then downsampling is a separable operation across the rows and columns.

Let us assume that $\boldsymbol{\Phi}$ denotes a $q \times q$ Gaussian blurring kernel and $\underline{\boldsymbol{Y}}_S(:,:,k) \in \mathbb{R}^{I_M \times J_M}$ be the matrix representation of the super-resolution image at the $k$th band. Then the convolution operation of image $\underline{\boldsymbol{Y}}_S(:,:,k)$ with the kernel $\boldsymbol{\Phi}$ can be modeled as:

$$\underline{\boldsymbol{Z}}_H(i,j,k) = \sum_{m=1}^{q} \sum_{n=1}^{q} \boldsymbol{\Phi}(m,n) \underline{\boldsymbol{Y}}_S\left(i-m', j-n', k\right), \quad (22)$$

where $m' = m - \lceil \frac{q}{2} \rceil$ and $n' = n - \lceil \frac{q}{2} \rceil$. Here, we have $\boldsymbol{\Phi}(m,n) = (1/2\pi\sigma^2) e^{-\frac{m'^2 + n'^2}{2}}$, which can be written as $\boldsymbol{\Phi}(m,n) = \phi(m)\phi(n)$, where $\phi(m) = (1/\sqrt{2\pi\sigma^2}) e^{-\frac{m'^2}{2}}$. Then, Eq. (22) takes the form

$$\boldsymbol{Z}_H(i,j,k) = \sum_{m=1}^{q} \sum_{n=1}^{q} \phi(m)\phi(n) \underline{\boldsymbol{Y}}_S\left(i-m', j-n', k\right) \quad (23)$$

which is a separable 2D convolution operation. Consequently, the blurring processing can be re-written as

$$\underline{\boldsymbol{Z}}_H(:,:,k) = \mathcal{T}_I(\boldsymbol{\phi}) \underline{\boldsymbol{Y}}_S(:,:,k) (\mathcal{T}_J(\boldsymbol{\phi}))^T,$$

where $\boldsymbol{\phi} = [\phi(1), \ldots, \phi(q)]^T$ and $\mathcal{T}_l(\boldsymbol{\phi})$ is the Toeplitz matrix that models the 1-D convolution operation of a vector $\boldsymbol{\phi}$ with a vector of size $l$ as a matrix vector multiplication.

The second step of the popular spatial degradation model is to downsample the blurred image by a factor of $d = d_1 d_2$. The 2-D downsampling operation of the blurred image $\boldsymbol{Z}_H$ can be cast as follows:

$$\underline{\boldsymbol{Y}}_H(i,j,k) = \sum_{m=1}^{I} \sum_{n=1}^{J} \boldsymbol{\delta}(m - id_1, n - jd_2) \underline{\boldsymbol{Z}}_H(m,n,k), \quad (24)$$

where $\boldsymbol{\delta}$ is the 2-d Kronecker Delta function. Using the separability property of the 2-D Kronecker Delta (i.e., $\delta(i,j) = \delta(i)\delta(j)$, where $\delta(i)$ is the 1-D Delta function), the transformation from $\underline{\boldsymbol{Y}}_S$ to $\underline{\boldsymbol{Y}}_H$ can be finally modeled as:

$$\underline{\boldsymbol{Y}}_H(:,:,k) = \boldsymbol{S}_1 \underline{\boldsymbol{Z}}_H(:,:,k) \boldsymbol{S}_2^T = \boldsymbol{P}_1 \underline{\boldsymbol{Y}}_S(:,:,k) \boldsymbol{P}_2^T \quad (25)$$

where $\boldsymbol{S}_1, \boldsymbol{S}_2$ are matrices that perform regular sampling of rows and columns respectively (they systematically choose 1 out of $d_1$ rows and 1 out of $d_2$ columns of $\underline{\boldsymbol{Z}}_H(:,:,k)$) and $\boldsymbol{P}_1 = \boldsymbol{S}_1 \mathcal{T}_I(\boldsymbol{\phi}), \boldsymbol{P}_2 = \boldsymbol{S}_2 \mathcal{T}_J(\boldsymbol{\phi})$.

We should mention that although we only showed the Gaussian blurring kernel case here, our tensor mode product based degradation model is compatible with any blurring kernel that factors to row and column blurring operators.

## APPENDIX D
## INITIALIZATION ALGORITHMS

In this section, we describe the algorithms that we propose to initialize the proposed STEREO and blind STEREO. The initialization approach computes factors $\boldsymbol{A}$ and $\boldsymbol{B}$ by the rank-$F$ CPD of $\underline{\boldsymbol{Y}}_M$. In the case where the downsampling operator is known, $\tilde{\boldsymbol{A}}$ and $\tilde{\boldsymbol{B}}$ are obtained as $\tilde{\boldsymbol{A}} = \boldsymbol{P}_1 \boldsymbol{A}$ and $\tilde{\boldsymbol{B}} = \boldsymbol{P}_2 \boldsymbol{B}$. Finally factor $\boldsymbol{C}$ is derived as solution of the linear equation $\boldsymbol{Y}_H = (\tilde{\boldsymbol{B}} \odot \tilde{\boldsymbol{A}}) \boldsymbol{C}^T$. In case where $\boldsymbol{P}_H$ is unknown, $\tilde{\boldsymbol{A}}$ and $\tilde{\boldsymbol{B}}$ are approximated by averaging out $d = \frac{I_M}{I_H}$ column entries of $\boldsymbol{A}$ and $\boldsymbol{B}$, respectively, to roughly mimic the blurring and downsampling process. Then matrix $\boldsymbol{C}$ can then be obtained as before. This procedure is illustrated in Algorithm 3.

---

**Algorithm 3:** Blind TenRec

**Initialization**: $F$
$\boldsymbol{A}, \boldsymbol{B}, \tilde{\boldsymbol{C}} \leftarrow \text{CPD}(\underline{\boldsymbol{Y}}_M)$
$\tilde{\boldsymbol{A}}(i,:) \leftarrow \sum_{k=d(i-1)+1}^{di} \boldsymbol{A}(k,:)$
$\tilde{\boldsymbol{B}}(i,:) \leftarrow \sum_{k=d(i-1)+1}^{di} \boldsymbol{B}(k,:)$
$\boldsymbol{C} \leftarrow \text{solve } \boldsymbol{Y}_H = (\tilde{\boldsymbol{B}} \odot \tilde{\boldsymbol{A}}) \boldsymbol{C}^T$

---